\documentclass[reprint,
 amsmath,amssymb,
 aps,
]{revtex4-1}
\usepackage{graphicx}
\usepackage{dcolumn}
\usepackage{bm}

\usepackage{siunitx}
\usepackage[disable]{todonotes} 

\begin{document}

\preprint{APS/123-QED}

\title{Calculation of a fluctuating entropic force by phase space sampling}

\author{James T. Waters}
\author{Harold D. Kim}%
 \email{harold.kim@physics.gatech.edu}
\affiliation{%
 School of Physics, Georgia Institute of Technology\\
 832 State Street, Atlanta, GA 30332-0430
}%


\date{\today}

\begin{abstract}

A polymer chain pinned in space exerts a fluctuating force on the pin point in thermal equilibrium. The average of such fluctuating force is well understood from statistical mechanics as an entropic force, but little is known about the underlying force distribution. Here, we introduce 
two phase space sampling methods that can produce the equilibrium distribution 
of instantaneous forces exerted by a terminally pinned polymer. In these 
methods, both the positions and momenta of mass points representing a freely 
jointed chain are perturbed in accordance with the spatial constraints and the
Boltzmann distribution of total energy. The constraint force for each 
conformation and momentum is calculated using Lagrangian dynamics. Using terminally pinned chains in space and on a surface, we show that the force distribution is highly asymmetric with both tensile and compressive forces. Most importantly, the mean of the distribution, which is equal to the entropic force, is not the most probable force even for long chains. Our work provides insights into the mechanistic origin of entropic forces, and an efficient computational tool for unbiased sampling of the phase space of a constrained system. 

\end{abstract}

\maketitle

\listoftodos
\onecolumngrid
\section{Introduction}
According to the second law of thermodynamics, a system tends to seek a higher
entropy state. When the increase in entropy can be gained through a change in 
a spatial coordinate, an effective force emerges. This force, purely entropic 
in origin, is a universal phenomenon, not specific to the underlying microscopic
Hamiltonian of the system. Examples of entropic force are found in all realms of
physics such as the Casimir force~\cite{kardar1999friction}, elastic force, and
depletion force~\cite{leckband2001intermolecular}. The concept of entropic force 
has even been applied to gravity
~\cite{jacobson1995thermodynamics,verlinde2011origin}.  

Polymers provide an excellent model system to study the entropic force both
theoretically and experimentally. They can be attached to a surface, confined 
in a volume, or constrained in a particular shape. Compared to polymers free 
in solution, geometrically confined polymers have reduced entropy. This
reduction in entropy leads to an entropic force, which may influence the rate of
detachment from a surface or the escape time of a polymer from confinement. The
confinement-mediated force was measured for a chain in a
tube~\cite{turner2002confinement}, and calculated using semianalytical means and
brownian dynamics simulations~\cite{prinsen2009force}. A simple heuristic
argument estimates its magnitude to be $\sim k_B T / a$, where $a$ is the
monomer length.

More involved calculations include computation of the full partition function
under confinement and its derivative with respect to
displacement~\cite{zhao2011gaussian}. Alternatively, this force can be calculated
from the local concentration of mass points near the surface
~\cite{guo2005rigid,odenheimer2005forces}. For a chain tethered to a hard wall,
the force increases with chain length, but saturates at $\sim k_B T / a$. For a
temperature \SI{310}{\kelvin} and a monomer length of \SI{3.4}{\angstrom}, this
corresponds to \SI{12.6}{\pico\newton}. This is consistent with
\SI{13.3}{\pico\newton} value for the pressure integrated over the entire
surface found by Bickel \emph{et al}.\cite{bickel2001local}, where a shorter monomer
length of \SI{3}{\angstrom} is assumed.

These methods are adequate to calculate the entropic force as the mean
thermodynamic force conjugate to displacement. However, entropic force itself 
is induced by a fluctuating quantity~\cite{bartolo2002fluctuations}. Several
examples in biology point to the importance of rare events that deviate from 
the average behavior~\cite{jeppesen2001impact,guo2011role}. Likewise, force
fluctuations may have a significant impact on chemical and biological
processes~\cite{koslover2012force}. Therefore, finding the full distribution of
instantaneous forces or its higher moments can give insights not easily
accessible by statistical mechanics.  

\begin{figure}
\begin{minipage}[c][\textheight]{\textwidth}
\includegraphics[width=8.25cm]{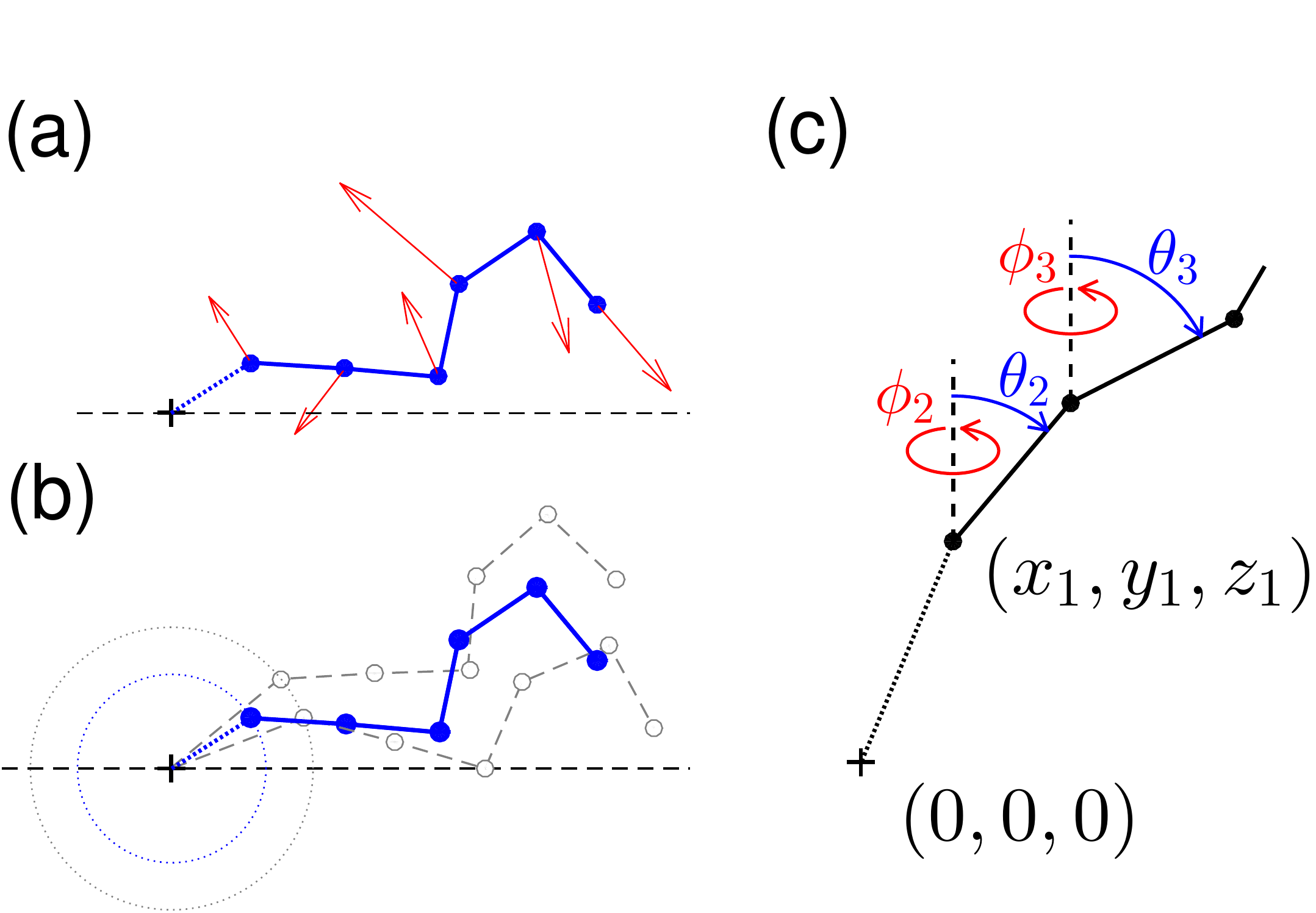}
\caption{The origin of force by a constrained chain. A freely jointed chain
(blue balls and sticks) is terminally pinned to a point (crosshair) and can 
move in full space or half space (dashed horizontal line). (a) Constraint
(centrifugal) force. When freely jointed mass points of the chain move with 
some velocities (red arrows), the pin point will experience a force. The mass
points can also exert force on the pin point when they recoil from the wall. 
(b) Thermodynamic (entropic) force. If the chain is pinned through a longer
tether, more conformations become available. For example, the first mass point
can explore a larger surface (compare blue and gray circles). Some 
conformations prohibited by the wall boundary can become acceptable as well 
when the tether is lengthened. This increase in conformational space leads to 
an increase in entropy, and thus a net force away from the surface. (c)
Generalized coordinates in the global frame. The first three mass points of the chain are shown 
as black balls. The first three Cartesian coordinates $(x_1, y_1, z_1)$
define the position of the first mass point with respect to the origin. The
tether length is $r_1$. The positions of the rest are defined by global 
zenith ($\theta$) and azimuthal ($\phi$) angles.}
\label{fig:schematic}
\end{minipage}
\end{figure}

In this study, we obtain the equilibrium force distribution due to a flexible
polymer pinned in space or to a surface (Fig.~\ref{fig:schematic}). The key 
idea of our approach is to make random moves in the phase space of generalized
position coordinates and their conjugate momenta and calculate microscopic
forces exerted by the system. This approach is different from conventional
Markov chain Monte Carlo methods that typically apply moves only in the 
position space. Because of the universality of entropic force, we expect our
method to be applicable to a wide range of dynamic systems.  

\section{Methods}
\subsection{Entropic force prediction}
The entropic force can be found from a thermodynamic calculation
\begin{equation}
F = \frac{\partial A}{\partial r} = -k_BT\frac{\partial \log Q}{\partial r}
\label{eq:force}
\end{equation}
where $A$ is the free energy, $Q$ is the partition function, and $r$ is the
generalized coordinate of interest. By measuring the change in the partition
function as the chain is slightly displaced from the pin point, we can compute
the entropic force. As shown in Fig.~\ref{fig:schematic}(\lowercase{b}), when the tether 
is lengthened, additional conformations become available. For the full space
case, the number of additional conformations is simply proportional to the 
area available to the first mass point. For the half-space case, we count 
the fraction of chain conformations compatible with the boundary
condition~\cite{waters2013equilibrium}, and calculate the small increase 
($\Delta Q$) upon varying $r_1$ by $\Delta r_1$ to numerically approximate
Eq.~\ref{eq:force}. 

\subsection{Phase space sampling}

\subsubsection{Monte Carlo sampling}

Beginning from some initial conformation and initial generalized momenta, we 
can generate a set of conformations for a coarse-grained chain in a heat bath.
The Hamiltonian of the chain in terms of generalized positions ($\mathbf{q}$)
and momenta ($\mathbf{p}$) will be given by
\begin{equation}
H(\mathbf{q},\mathbf{p}) = U(\mathbf{q}) + \frac{1}{2}\mathbf{p}^\mathsf{T}
\mathbf{M}^{-1}\mathbf{p}
\end{equation}
where the first term $U$ is a potential energy based on chain deformation, 
and the second term is the kinetic energy expressed with the mass metric
$\mathbf{M}$. For a freely jointed chain (FJC), $U$ is zero. We use the FJC model in this study because we want to obtain forces purely entropic of origin, but not due to physical interactions. The FJC model effectively coarse-grains an arbitrary polymer into Kuhn-length segments. The mass metric 
can be found from the Jacobian
\begin{equation}
(\mathbf{M})_{ij}=m(\mathbf{J}^{\mathsf{T}}\mathbf{J})_{ij}=m \frac{\partial
x_k}{\partial q_i}\frac{\partial x_k}{\partial q_j}
\end{equation}
$\mathbf{M}$ depends on $\mathbf{q}$ in non-Cartesian space. We choose $\theta$
and $\phi$ angles of each link measured in the global reference frame as the
generalized coordinates (Fig. \ref{fig:schematic}(c)). These are chosen over
internal coordinates, as a change to the orientation of one link in the global
frame will only affect two columns of the Jacobian matrix.

At each step, one of the momentum coordinates can be perturbed, and the new
kinetic energy computed. The trial step can be accepted or rejected according 
to the Metropolis criterion on the total energy. The position coordinates can
also be perturbed, which will require computing the change in kinetic energy
(via a change in $\mathbf{M}(\mathbf{q})$) before the move can be accepted or
rejected.

\subsubsection{Equilibrium sampling}

For longer chains, we sample points in the phase space using modal velocity
decomposition~\cite{jain_equipartition_2012}. The key idea is that microstates
specified by both $\mathbf{q}$ and $\mathbf{p}$ can be directly sampled from 
the probability density function ($\rho(\mathbf{q},\mathbf{p})$) to build the
distribution of any microscopic variable. Our method is different from methods
categorized as equilibrium sampling~\cite{zuckerman2011equilibrium} in that we
sample both positions and momenta. For a canonical ensemble,
$\rho(\mathbf{q},\mathbf{p})$ is given by
\begin{equation}
\rho(\mathbf{q},\mathbf{p}) d^{2N}qd^{2N}p \propto
e^{-\frac{\beta}{2}\dot{\mathbf{q}}^\mathsf{T}
\mathbf{M(\mathbf{q})}\dot{\mathbf{q}}} d^{2N}qd^{2N}p ,
\end{equation}
where $\beta=1/k_BT$. In this form, sampling individual variables $q_i$ and $\dot{q}_i$ is not
straightforward because they are coupled through $\mathbf{M(\mathbf{q})}$.
Matrix diagonalization can be used to separate variables, but is not
computationally efficient. Instead, the Cholesky decomposition can be used to
express the symmetric, positive mass metric as
\begin{equation}
\mathbf{M} = m\boldsymbol{\mu}\boldsymbol{\mu}^\mathsf{T}
\end{equation}
where $\mathbf{\mu}$ is a lower triangular matrix. By defining modal velocities
as $\mathbf{v}=\boldsymbol{\mu}^\mathsf{T}\dot{\mathbf{q}}$, we can express the
probability in a more separable form
\begin{equation}
\rho(\mathbf{q},\mathbf{p}) d^{2N}qd^{2N}p \propto
\sqrt{\text{det}(\mathbf{M(\mathbf{q})})} d^{2N}q\cdot
e^{-\frac{\beta}{2}\mathbf{v}^\mathsf{T}\mathbf{v}}d^{2N}v
\end{equation}
In this form, modal velocities ($v_i$) can be sampled independently of each
other from a Gaussian distribution. However, positions ($q_i$) become coupled
through $\text{det}(\mathbf{M(\mathbf{q})})$, which is known as the Fixman
correction~\cite{fixman1974classical}. Hence, we adopt a hybrid approach where 
we first sample a microstate using normally distributed modal velocities and
weight the microstate-dependent variable by the Fixman term. Calculating the
determinant of the dense metric tensor is computationally intensive, however
Fixman demonstrated that this is equivalent to the determinant of a smaller,
tri-diagonal metric of the constrained coordinates (for details, see
Supplemental Information).

\subsection{Computing constraint force from positions and momenta}
We consider a flexible chain terminally pinned in space or to a surface. We can
use Lagrangian dynamics to calculate the constraint force as a function of
positions and velocities of the $N$ mass points. Because distances between all
adjacent mass points are constant, there are $2N+1$ degrees of freedom. We use
three Cartesian coordinates $(x_1,y_1,z_1)$ for the first mass point, and
$2(N-1)$ angles ($\theta$ and $\phi$) for all other mass points of the chain
(Fig. \ref{fig:schematic} (c)). Choosing generalized coordinates this way
simplifies subsequent calculations. The Lagrangian is given by
\begin{equation}
L = \frac{1}{2}\dot{\mathbf{q}}^\mathsf{T} \mathbf{M} \dot{\mathbf{q}}
\end{equation}
For the first mass point of the chain to be radially constrained at distance
$r$, we introduce a Lagrange multiplier $\lambda$
\begin{equation}
L^{\prime} = \frac{1}{2}\dot{\mathbf{q}}^\mathsf{T} \mathbf{M}
\dot{\mathbf{q}}+\lambda (r_1-r)
\end{equation}
where $r_1=\sqrt{x_1^2+y_1^2+z_1^2}$. Using $L^{\prime}$ in the Euler-Lagrange
equation with respect to $q_{k=1,2,3}$
\begin{equation}
\frac{d}{dt} \left(M_{ki} \dot q_i \right) = \frac{1}{2}\frac{\partial
M_{ij}}{\partial q_k}\dot q_i \dot q_j + \lambda\frac{\partial r_1}{\partial
q_k}
\end{equation}
The first term on the right hand side vanishes as $\mathbf{M}$ does not depend
on the Cartesian coordinates of the first mass point. Carrying out the
differentiation on the left hand side produces an expression
\begin{equation}
M_{ki} \ddot q_i + \frac{\partial M_{ki}}{\partial q_j} \dot q_i \dot q_j = 
\lambda\frac{q_k}{r_1}
\end{equation}
The second derivative of these terms must be solved for using the equation of
motion for the angular coordinates.
\begin{equation}
M_{ij} \ddot q_j + \frac{\partial M_{ij}}{\partial q_k} \dot q_j \dot q_k = 
\frac{1}{2}\frac{\partial M_{jk}}{\partial q_i}\dot q_j \dot q_k
\end{equation}
Solving for $\ddot q_j $ gives us
\begin{equation}
\ddot q_j =  M_{ij}^{-1} \left(- \dot q_l \dot q_k \frac{\partial
M_{il}}{\partial q_k} +  \frac{1}{2}\dot q_l \dot q_k \frac{\partial
M_{lk}}{\partial q_i} \right)
\end{equation}
We can express this more concisely as the geodesic equation using the
Christoffel connection coefficient, and making use of the symmetry in the first
term under exchange of $l$ and $k$
\begin{equation}
\ddot q_j =  -\dot q_l \dot q_k \Gamma_{lk}^j
\end{equation}
where 
\begin{equation}
 \Gamma^i_{jk} = \frac{1}{2}M_{il}^{-1} \left(\frac{\partial M_{jl}}{\partial
 q_k} + 
 \frac{\partial M_{kl}}{\partial q_j} -
 \frac{\partial M_{jk}}{\partial q_l} \right)
\end{equation}
Substituting this back in produces the Cartesian components of the constraint
force
\begin{equation}
\lambda\frac{q_k}{r_1} = 
\dot q_i \dot q_j \left(\frac{\partial M_{ik}}{\partial q_j} - \Gamma^l_{ij}
M_{lk} \right) = 
\dot q_i \dot q_j \nabla_i M_{jk}
\label{eq:lambda}
\end{equation}
where $\nabla_i$ is the covariant derivative in the constrained subspace.

\section{Results}

\subsection{A single point constraint in space}
\begin{figure}
\begin{minipage}[c][\textheight]{\textwidth}
\includegraphics[width=17cm]{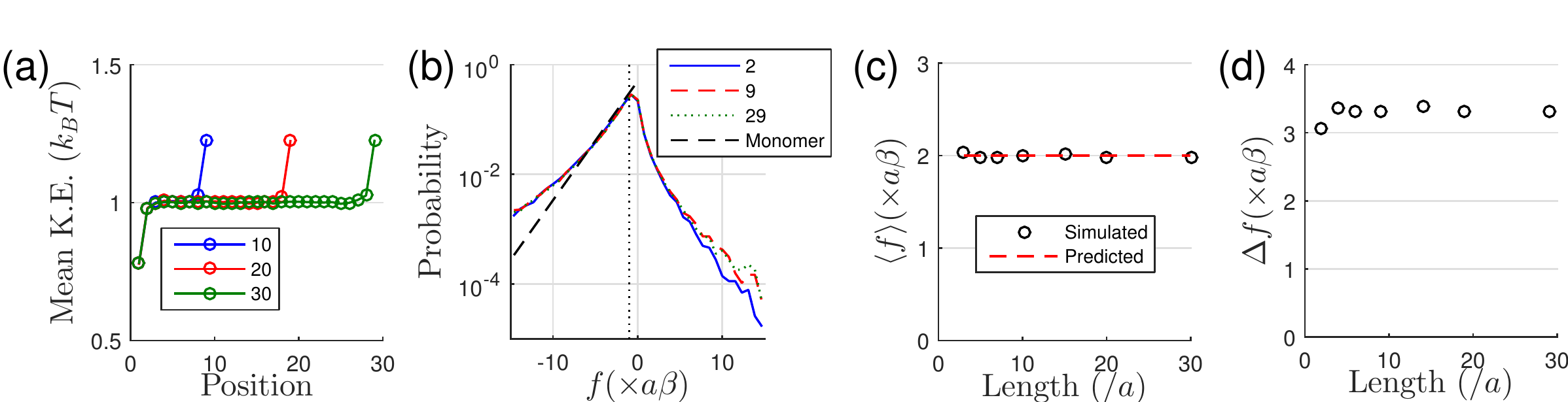}
\caption{Chain pinned in full space. (a) Distribution of kinetic energy at
different points along the chain. The first mass point after the pinning
position has less energy than average, and the last one has slightly more. The
mean kinetic energy remains  $N k_BT$ in all cases. Shown are results from
different chain lengths ($N$=10, 20, and 30). (\lowercase{b}) Distribution of radial
constraint forces. While the mean constraint force equals that due to a single
mass point, additional links in the polymer chain increase likelihood of strong
pulling forces and even introduce the possibility of compressive forces. The most probable force is near $-1$ shown by the vertical dotted line. Distributions are computed for different chain lengths ($N$=2, 9, and 29) from 150 K samples. (c)
Mean constraint force and entropic force. The average force (black circles) is
independent of length, and matches the predicted entropic force (red line),
which results from the first mass point confined to the surface of a sphere
moving to a larger radius. (d) Standard deviation of constraint forces. The
standard deviation is on the same order of magnitude as the mean force, and
quickly saturates as chain length increases.}
\label{fig:fullspace}
\end{minipage}
\end{figure}

We first considerd a thermally equilibrated flexible chain with one end pinned
to a single point in full three dimensional space. By using the FJC model without a potential term, we can focus solely on entropic contributions to the force. We used the Monte Carlo phase
space sampling method to build a canonical ensemble of chains. As shown in
Fig.~\ref{fig:fullspace}(a), kinetic energy is not equally partitioned between
the mass points of the chain due to the constraints. The mass point nearest to
the pin point has $\sim25$\% less kinetic energy whereas the endmost mass point
has $\sim25$\% more. Most mass points in the middle have $k_B T$ on average,
which would be expected for two degrees of freedom. Using these chain
conformations and momenta, we calculated individual constraint forces from
Lagrangian mechanics (Eq.~\ref{eq:lambda}). The distribution of constraint
forces is highly asymmetric, including both positive and negative values with a
sharp cusp. The mean constraint force is a negative pulling force, and does not
depend on the chain length. Due to high skewness of the distribution, the mean
force ($\sim 2 k_BT/a$) is larger in magnitude than the most probable force ($\sim 1 k_BT/a$).
The standard deviation also has little dependence on the chain length (Fig.~\ref{fig:fullspace}(d)). 

We compared this mean constraint force with the entropic force ($F$) predicted
by statistical mechanics. To seek more conformations, i.e., to increase entropy,
the chain would prefer detachment from the pin point. As a result, the chain
exerts a radial entropic force on the pin point. If the distance ($r$) between
the pin point and the first mass point increases by $\Delta r$, the number of
conformations will increase in proportion to the spherical surface spanned by
the mass point. Therefore, the partition function ($Q$) is proportional to $4\pi
r^2$, and the force at monomer distance $a$ is given by Eq.~\ref{eq:force}
\begin{equation}
F \approx \frac{k_B T}{Q}\frac{Q(a+\Delta r)-Q(a)}{\Delta r} = \frac{2k_BT}{a}
\label{eq:entropicforce}
\end{equation}
This entropic force is identical to the mean constraint force in magnitude, and
is also independent of chain length as shown in Fig.~\ref{fig:fullspace}(c).

Interestingly, the mean constraint force due to the entire chain is equal to the
mean centrifugal force exerted by a single mass point. A radially constrained
mass point has only two degrees of freedom, and therefore, according to the
equipartition theorem, its mean kinetic energy is equal to $k_B T$. The mean of
the centrifugal force ($f_c$) is 
\begin{equation}
\langle f_c \rangle = \left\langle \frac{mv^2}{a} \right\rangle =
\frac{2k_BT}{a}
\end{equation}
Since the energy ($E=mv^2/2$) of a canonical ensemble is Boltzmann-distributed,
the centrifugal force is exponentially distributed. 
\begin{equation}
p(E)dE = e^{-\beta E}dE = \frac{a}{2}e^{-\beta a f_c/2}df_c
\label{eq:boltzmann}
\end{equation}
As shown in Fig.~\ref{fig:fullspace}(\lowercase{b}), this exponential distribution (dashed
line) can explain much of the distribution of negative constraint forces.
However, positive forces and a significant fraction of large negative forces
deviate from this distribution. This result suggests that freely jointed mass
points beyond the nearest one can have a collective impact on the pin point. 

\begin{figure}
\begin{minipage}[c][\textheight]{\textwidth}
\includegraphics[width=8.25cm]{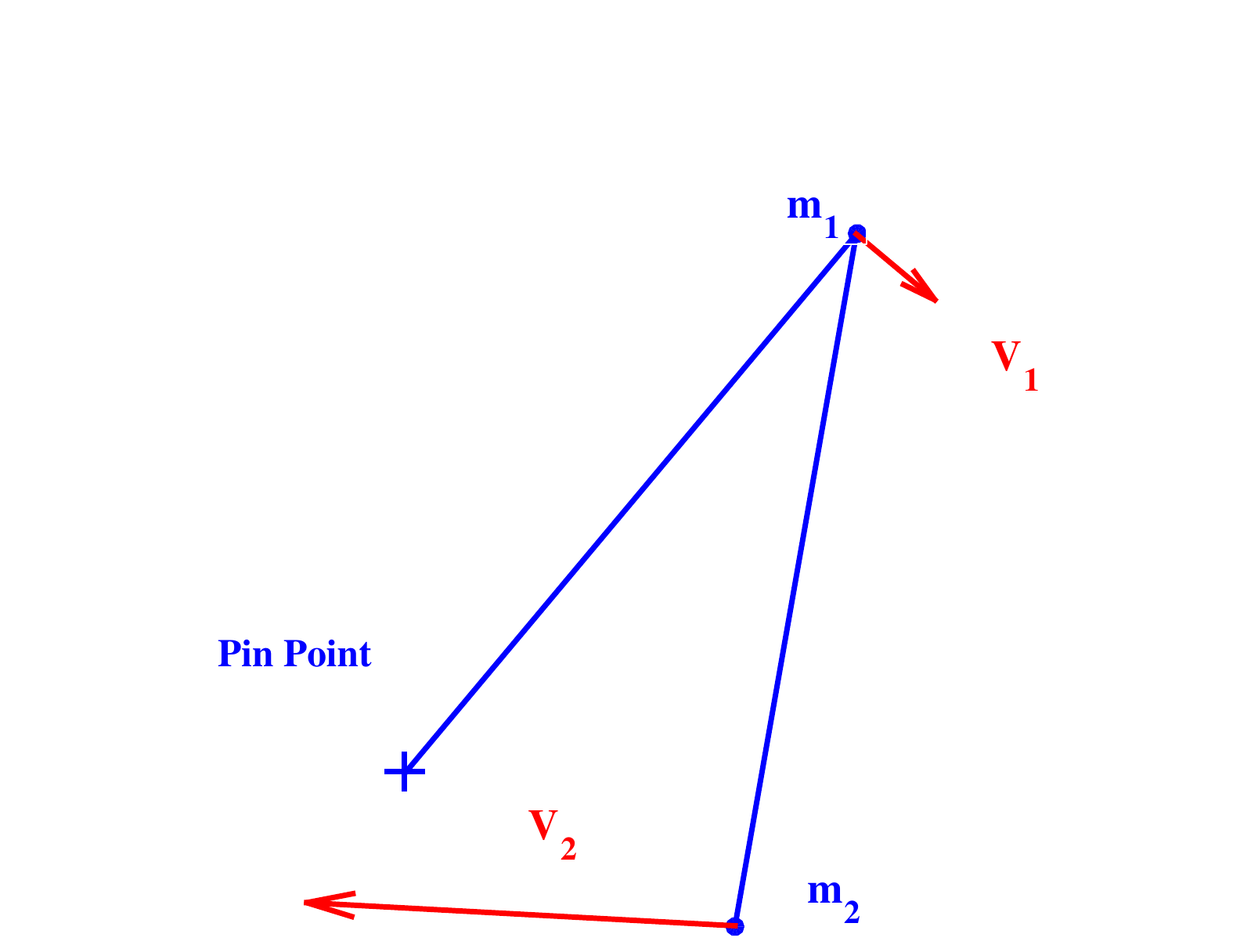}
\caption{Origin of compressive forces. Some fraction of conformations (blue) and
velocities (red) produce compressive forces along the first link in the chain,
in contrast to the expectation from a single link acting as a pendulum. These
occur when the chain is sharply bent, and the downstream mass points are moving
more rapidly than those closer to the pin point ($v_2 > v_1$).}
\label{fig:rad_positive}
\end{minipage}
\end{figure}

To understand combined force generation by freely jointed mass points, we
examined conformations that produce anomalous force values outside the range
attributable to a single monomer. As shown in Fig.~\ref{fig:rad_positive},
instances where the force has the positive sign result from sharp bending and
large velocities. For the simple case of a trimer in two dimensions, we can
derive a condition for compressive force
\begin{equation}
 \dot \theta_2^2 \cos(\theta_2-\theta_1) > 2\dot\theta_1^2
\end{equation}
This corresponds to the case where the centrifugal force exerted by $m_2$ on
$m_1$ towards the pin point is more than twice that exerted by $m_1$ on the pin
point in the opposite direction.

\subsection{A single point constraint on a plane}
\begin{figure}
\begin{minipage}[c][\textheight]{\textwidth}
\includegraphics[width=17cm]{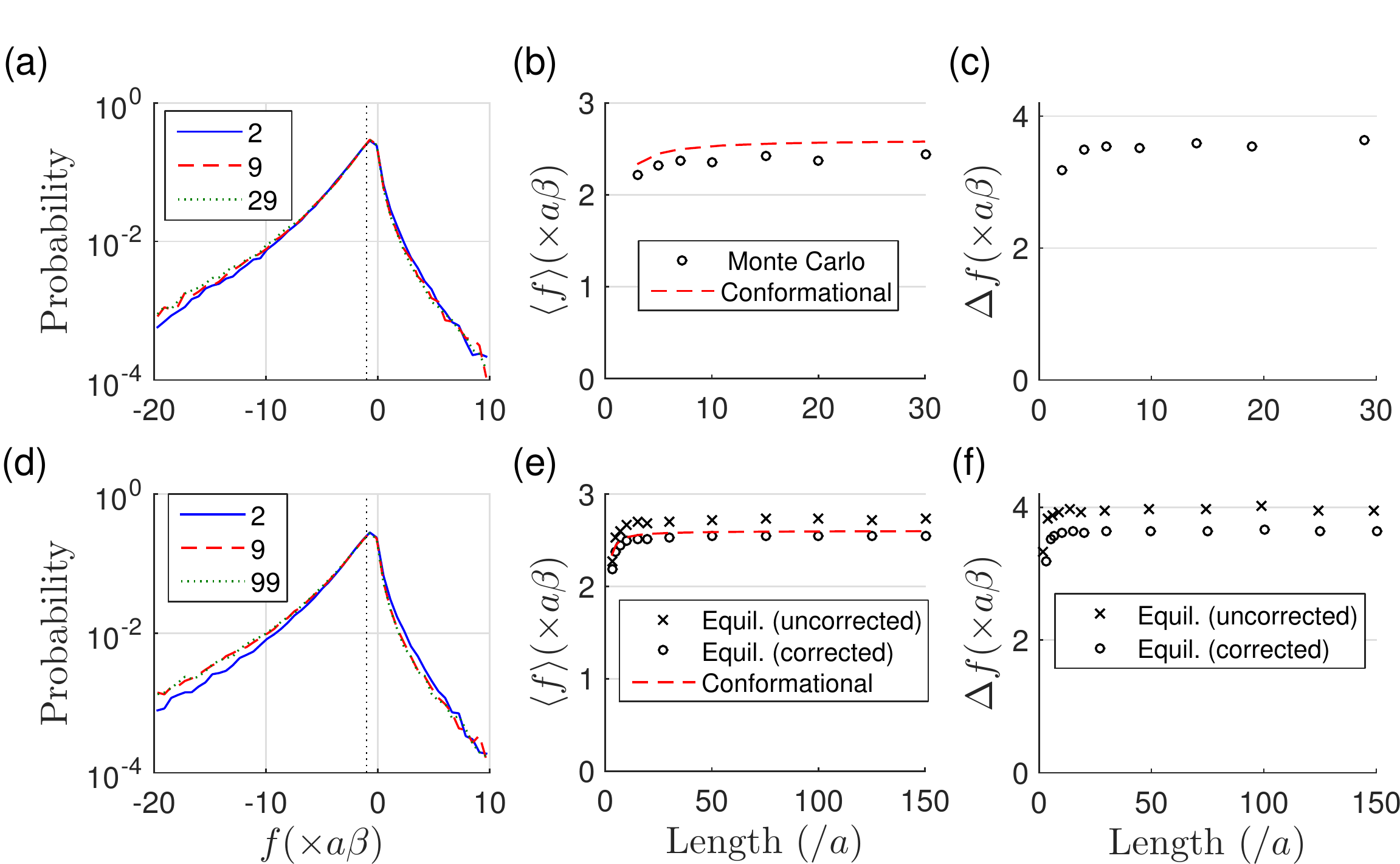}
\caption{Chain pinned to plane. (a) Distribution of forces for phase-space
sampled chains. The surface effect shifts the mean of the distribution slighty,
but the shape is not changed relative to the free case. The most probable force is near $-1$ shown by the vertical dotted line. Distributions are
computed for different chain lengths ($N$=2, 9, and 29), from 150 K samples. (b) Mean force from
chains sampled via phase space Monte Carlo (circles) and thermodynamic force
from conformation space partition function (dashed line). The average force lies
above the $2/a\beta$ value of the free chain case, as additional entropy is
gained by pulling away from the surface. The mean force exhbits some length
dependence not seen in the free chain case, however it quickly saturates around
10 links in the polymer chain. (c) Standard deviation of sampled forces. As in
the free chain case, the deviation is on the same order of magnitude as the mean
force. It grows slightly as a function of length, quickly saturating around 5-7
links.  (d) Distribution of forces for conformational space sampled chains. The
same distribution observed in (a) persists as we consider longer chains
($N$=2, 9, and 99). The most probable force is near $-1$ shown by the vertical dotted line. The curves in the figure represent 300 K samples. (e) Mean force from equilibrium sampled chains, before
applying the Fixman correction ($\times$) and after ($\circ$), displayed along side the
partition function prediction (dashed line). The mean force remains constant as
we explore larger lengths than were accessible in (b). (f) Standard deviation
for force from equilibrium sampled chains, with ($\circ$) and without ($\times$) the
Fixman correction. }
\label{fig:force_kinetic}
\end{minipage}
\end{figure}

We next considered a flexible chain attached to a surface. This confinement
geometry arises in biology
problems~\cite{petrov2013effects,ludington2015systematic} and polymer
applications~\cite{milchev2010ejection}. The phase space sampling method applies
the same way except that we only accept chain conformations that do not cross
the boundary. We computed constraint forces at different chain lengths. 
As shown in Fig.~\ref{fig:force_kinetic}(a) and (d), the distribution of constraint forces
is highly asymmetric, similar to the full space case. Most forces are exerted in
the outward direction, but a significant fraction of forces are compressive. The
mean constraint force increases slightly as a function of chain length up to ten
monomer lengths, but quickly reaches a constant value. The mean constraint force
plateaus at about $2.5k_BT/a$, which is \SI{25}{\percent} higher than the full
space case. 

For longer chains ($N>50$), phase space sampling becomes computationally
expensive. We therefore adopted an equilibrium sampling method to calculate
constraint forces. This method requires weighting by the Fixman correction. We
computed a raw force distribution (uncorrected, `$\times$' in
Fig.~\ref{fig:force_kinetic}), and later applied this correction (`$\circ$' in
Fig.~\ref{fig:force_kinetic}). Applying the Fixman correction lowers the average
value of the force, as the largest forces are correlated with the largest
extensions, which have a low weighting factor. This equilibrium method after
correction produces similar distributions to the Monte Carlo method, highly
asymmetric with more frequent negative than positive forces. 

To compute the corresponding entropic force, we consider the increase in the
accepted chain conformations upon increasing the distance to the first mass
point. The partition function can then be numerically differentiated using
Eq.~\ref{eq:entropicforce} to obtain the entropic force. The result did not
depend on the size of displacement chosen. As shown by the red dashed line in
Fig.~\ref{fig:force_kinetic}(\lowercase{b}) and (e), the entropic forces based on
conformational space only are slightly larger than the mean constraint forces
($2.6 k_BT/a$ vs. $2.5 k_BT/a$). 

\section{Discussion}
We computed constraint forces exerted on a FJC in full space or half space by
classical and statistical mechanical methods. Statistical mechanics was used to
describe thermal fluctuations of the chain microstate, and classical mechanics
was used to calculate the force required to constrain the chain against these
thermal fluctuations. We found that the mean of constraint forces is equal to
the entropic force calculated from the spatial derivative of the free energy.
However, because constraint forces are asymmetrically distributed, the mean
force is not equal to the most probable force. The force distribution is largely
independent of the chain length beyond ten monomer lengths.

In this study, we treat the FJC as a canonical ensemble, which exchanges heat
with the environment at a constant temperature. Our equilibration of microstates
in the phase space is similar to the Andersen
thermostat~\cite{andersen1980molecular} used in molecular dynamics with the added
capability of simultaneously satisfying geometric constraints. At any instant,
the mass points of the chain move with certain velocities at certain positions
under the constraints set by the FJC model and the boundary. At a different time
point, the mass points will adopt a different conformation with a different set
of velocities as a result from collisions with the surrounding molecules of the
heat bath. For each set of positions and momenta reset as a result of heat
exchange, we calculate the microscopic constraint force. 

The most nontrivial result from this study is that the force distribution is not Gaussian even in the long chain limit. Because the mass points are constrained in the freely jointed chain model, their impact on the tether are neither identical nor independent. As a result, the constraint force does not obey the central limit theorem. Consequently, the most probable value and the average value of force do not coincide with each other. Therefore, the entropic force that we derive from statistical mechanics might not be the most relevant quantity in all circumstances. 

The fluctuation or second cumulant of an extensive thermodynamic variable can be
easily calculated from the partition function when its conjugate intensive
variable is fixed~\cite{marconi2008fluctuation}. Textbook examples are energy
fluctuation in a canonical ensemble and particle number fluctuation in a grand
canonical ensemble. On the other hand, the fluctuation of an intensive variable
such as force cannot be derived from the partition function when the conjugate
extensive variable is fixed, i.e. in a fixed-distance
ensemble~\cite{rudoi2000thermodynamic,gerland2001force}. Here, we use the FJC
terminally constrained in full space to illustrate this point.

The constrained partition function ($Q_c$) is given by (See Supplemental
Information.) 
\begin{subequations}
\begin{align}
Q_c(r) &=\int e^{-\beta H_c(\mathbf{q},\mathbf{p})} d^{2N}qd^{2N}p \\
&=\int e^{-\frac{\beta}{2}\mathbf{p}^\mathsf{T} \mathbf{M}^{-1}\mathbf{p}}
d^{2N}qd^{2N}p ,
\label{eq:constrained}
\end{align}
\end{subequations}
where $H_c$ is the Hamiltonian of the constrained chain. The mass metric
$\mathbf{M}$ is a $2N\times2N$ matrix that depends on $2N$ global angles
$\mathbf{q}$. The dependence of the partition function on the distance between
the end and the pin point ($r$) is explicity shown. Entropic force ($F$) arises
because the constrained partition function increases with $r$, which is
reflected by the thermodynamic relation:
\begin{subequations}
\begin{align}
F &= \frac{1}{\beta}\frac{\partial \log Q_c(r)}{\partial r} \\
 &= \int \frac{1}{\beta}\frac{\partial e^{-\beta
 H_c(\mathbf{q},\mathbf{p})}}{\partial r} d^{2N}qd^{2N}p / Q_c(r)
 \label{eq:insert}\\
 &= -\int \frac{\partial H_c(\mathbf{q},\mathbf{p})}{\partial r} \frac{e^{-\beta
 H_c(\mathbf{q},\mathbf{p})}}{Q_c(r)} d^{2N}qd^{2N}p \\
 &= -\left\langle \frac{\partial H_c(\mathbf{q},\mathbf{p})}{\partial r}
 \right\rangle_c .
\end{align}
\end{subequations}
The quantity in the constrained ensemble average can be identified with a
microscopic force, which we denote as $f$. The second moment of $f$ can be shown
to be
\begin{equation}
\left\langle \Delta f^2 \right\rangle_c = \frac{1}{\beta^2}\frac{\partial^2 \log
Q_c(r)}{\partial r^2}+ \left\langle \frac{\partial^2H_c}{\partial r^2}
\right\rangle_c
\end{equation}
Whereas the first term can be calculated from the constrained partition
function, the second term depends on the microstate of the system. If the rigid
constraint is relaxed to a spring, the second term is simply the stiffness
($\kappa$) of the spring. In this case, force fluctuations can be easily
calculated, similar to those for a particle trapped in a harmonic
potential~\cite{manosas2005thermodynamic,mehraeen2012intrinsic}. However, the
force fluctuation diverges in the stiff limit
($\kappa\rightarrow\infty$)~\cite{gerland2003mechanically}. Therefore, to obtain
force fluctuations with a hard constraint, one has to resort to computational
means.  

We showed that for a terminally pinned chain in full space, our phase space
sampling method produces constraint forces whose mean is equal to the entropic
force $2k_BT/a$. Using Eq.~\ref{eq:constrained}, the entropic force can be
worked out analytically. Integrating out the momentum dependence of $Q_c(r)$,
one obtains
\begin{equation}
Q_c(r)=\text{const} \times \int \sqrt{\text{det}(\mathbf{M})} d^{2N}q
\label{eq:qc}
\end{equation}
$Q_c(r)$ depends on $r$ through the mass metric $\mathbf{M}$. It can be shown
that $\sqrt{\text{det}(\mathbf{M})} \propto r^2$ (Supplemental Information), and
therefore the entropic force is given by
\begin{equation}
F =-\frac{2k_B T}{r} .
\end{equation}

In comparison, the entropic force due to the pinned chain in a half space is
difficult to derive because the limits of integration imposed by the half plane
depend on $\mathbf{q}$ in some complex fashion. 

People have used the method of images to calculate a simpler partition function
in conformational space only~\cite{zhao2011gaussian,hammer2014ideal},
\begin{equation}
Q(h)\propto\int_\text{half space}
d^{2N}q\propto\text{erf}\left(h\sqrt{\frac{3}{2Na^2}}\right)
\label{eq:wrongpartition}
\end{equation}
where $h$ is the height of the pin point from the surface. By taking the
derivative of $\log Q(h)$ with respect to $h$, a vertical entropic force can be
derived. This vertical force increases with chain length for short chains, but
is nearly constant at $k_BT/h$ for long chains. This force, however, is expected
to be different from the mean constraint force obtained from our phase sampling
methods for several reasons. First, the method of images yields only the
vertical component of the entropic force out of mathematical convenience.
Second, the analytical expression (Eq.~\ref{eq:wrongpartition}) is correct only
in the Gaussian limit of long chains. Third, the conformational partition
function in Eq.~\ref{eq:wrongpartition} does not include the kinetic effect.
However, it is unclear if removing the kinetic effect would significantly alter
the mean force prediction. Therefore, we numerically computed the conformational
partition function as a function of $r$ instead of $h$, and obtained a
thermodynamic force conjugate to $r$ using Eq.~\ref{eq:entropicforce}. This
force is only slightly larger than the entropic force obtained by phase space
sampling as shown in Fig.~\ref{fig:force_kinetic}(b) and (e).

As the chain length becomes longer, the Monte Carlo method becomes
computationally expensive. A faster method is to sample conformations directly
from an a priori distribution. For example, to simulate a canonical ensemble of
ideal gas, one can randomly sample velocities from the Maxwell-Boltzmann
distribution and assign them to randomly positioned particles. Such equilibrium
sampling, however, is not straightforward for a constrained system due to the
coupling of $\mathbf{p}$ through $\mathbf{M}$. By decomposing $\mathbf{M}$, we
can obtain modal velocities that individually obey a Gaussian distribution. Such
a transformation skews the distribution by the Fixman term, which must be
corrected for to unbiasedly sample the phase space. Our results show that this
correction is warranted (Fig.~\ref{fig:force_kinetic}(e) and (f)).

Both phase space sampling methods we introduce in this work are based on
theoretically correct interpretations of the constrained partition function. The
Monte Carlo phase space sampling method has the advantage of not requiring the
Fixman correction to produce the correct average force prediciton. Additionally,
this method may prove more useful for geometries such as polymer chains with a
constant end-to-end distance. In such cases, selecting conformations that fit a
given distribution is difficult but perturbing a conformation that already lies
within that distribution is comparatively easy. The chief disadvantage of this
method is the computational time required to perturb the chain, and the large
number of steps required to adequately explore the conformational space. 

In contrast, the equilibrium phase space sampling, where uncorrelated
conformations are selected, and momenta assigned after, is much quicker for free
chains. This lack of correlation may more easily lend this method to large-scale
parallelization as well. One minor disadvantage of this method is that it
introduces a discrepancy in the mean force, unless the sampled conformations are
weighted by the Fixman correction. 

\section{Conclusions}
We introduced two different computational methods to sample microstates of a
freely jointed chain in an unbiased manner. Unlike mainstream Monte Carlo
methods that explore the conformational space only, our methods explore the
complete phase space to obtain full distribution of any velocity-dependent
microscopic variable in thermal equilibrium. We applied these methods to a
terminally pinned chain in full space and half space to calculate constraint
forces. The force distribution was non-Gaussian and asymmetric with both tensile
and compressive forces. Most notably, the most likely force was smaller in
magnitude than the mean force, which is more commonly considered both
theoretically and experimentally. The constraint force exhibited little
chain-length dependence beyond $\sim10$ monomer lengths. The presence of a plane
leads to a larger mean constraint force. This work can be extended to
investigate fluctuating forces using more realistic polymer models and
confinement. 

\section{Acknowledgement}
The authors acknowledge financial support from Georgia
Institute of Technology and the Burroughs Wellcome Fund
Career Award at the Scientific Interface. We thank Tung Le, Jiyoun Jeong, and the rest of the Kim lab
for careful reading of the manuscript. We also thank Dr. Kurt Wiesenfeld and Dr. Toan Nguyen for helpful discussions.

\bibliography{main}
\clearpage

\clearpage
\newcommand{\beginsupplement}{%
        \setcounter{table}{0}
        \renewcommand{\thetable}{S\arabic{table}}%
        \setcounter{figure}{0}
        \renewcommand{\thefigure}{\arabic{figure}}%
        \setcounter{equation}{0}
        \renewcommand{\theequation}{S\arabic{equation}}%
}

\beginsupplement
\section*{Supplemental Information}

\subsection{Dependence of the mass metric on the end constraint} 
The Cartesian coordinates of the n-th mass point are given by ($n \ge 2$). 
\begin{align*}
x_n &= x_1+\sum_{i=2}^N \sin \theta_i \cos \phi_i \\
y_n &= y_1+\sum_{i=2}^N \sin \theta_i \sin \phi_i \\
z_n &= z_1+\sum_{i=2}^N \cos \theta_i .
\end{align*}
It is assumed that all adjacent mass points are joined by an inextensible linker
of unit length. The spherical coordinates of the first mass point are given by 
\begin{align*}
x_1 &= r_1 \sin \theta_1 \cos \phi_1 \\
y_1 &= r_1 \sin \theta_1 \sin \phi_1 \\
z_1 &= r_1 \cos \theta_1 .
\end{align*}
We define two sets of generalized coordinates
$\mathbf{u}=\{r_1,\theta_i,\phi_i\}$ and a subset of unconstrained coordinates
$\mathbf{q}=\{\theta_i, \phi_i\}$ where $i=1,2,...,N$. The Jacobian matrix
$\mathbf{J}$ with respect to $\mathbf{u}$ can be constructed from smaller
Jacobians for individual mass points. The Jacobian matrix is a $3\times3$ matrix
for the first mass point:
\begin{equation*}
\mathbf{J}_1=
\begin{bmatrix}
\dfrac{\partial x_1}{\partial r_1} & \dfrac{\partial x_1}{\partial \theta_1} &
\dfrac{\partial x_1}{\partial \phi_1} \\[1em]
  \dfrac{\partial y_1}{\partial r_1} & \dfrac{\partial y_1}{\partial \theta_1} &
  \dfrac{\partial y_1}{\partial \phi_1} \\[1em]
  \dfrac{\partial z_1}{\partial r_1} & \dfrac{\partial z_1}{\partial \theta_1} &
  \dfrac{\partial z_1}{\partial \phi_1}\end{bmatrix}
= \begin{bmatrix}
      \sin \theta_1 \cos \phi_1 & r_1 \cos \theta_1 \cos \phi_1 & - r_1 \sin
      \theta_1 \sin \phi_1 \\
      \sin \theta_1 \sin \phi_1 & r_1 \cos \theta_1 \sin \phi_1 & r_1 \sin
      \theta_1 \cos \phi_1 \\
      \cos \theta_1 & - r_1 \sin \theta_1 & 0 \end{bmatrix},
\end{equation*}
and a $3\times2$ matrix for the rest:
\begin{equation*}
\mathbf{J}_{i\ge2}=
\begin{bmatrix}
\dfrac{\partial x_i}{\partial \theta_i} & \dfrac{\partial x_i}{\partial \phi_i}
\\[1em]
\dfrac{\partial y_i}{\partial \theta_i} & \dfrac{\partial y_i}{\partial \phi_i}
\\[1em]
\dfrac{\partial z_i}{\partial \theta_i} & \dfrac{\partial z_i}{\partial
\phi_i}\end{bmatrix}
= \begin{bmatrix}
\cos \theta_i \cos \phi_i & - \sin \theta_i \sin \phi_i \\
\cos \theta_i \sin \phi_i & \sin \theta_i \cos \phi_i \\
\sin \theta_i & 0 \end{bmatrix} .
\end{equation*}
The total Jacobian can be expressed as a block matrix: 
\begin{equation*}
\mathbf J = \begin{bmatrix} 
\mathbf{J}_{1} & 0 & \cdots & 0 \\ \mathbf{J}_{1} & \mathbf{J}_{2} & \cdots &  0
\\
\vdots & \vdots & \ddots & \vdots \\
\mathbf{J}_{1} & \mathbf{J}_{2} & \cdots & \mathbf{J}_{N} 
\end{bmatrix} .
\end{equation*}
The mass matric becomes a $(2N+1)\times(2N+1)$ square matrix:
\begin{equation*}
\mathbf{M}_u = m\mathbf{J}^\mathsf{T} \mathbf{J}=m\begin{bmatrix} 
N\mathbf{J}_1^\mathsf{T} \mathbf{J}_{1} & (N-1)\mathbf{J}_1^\mathsf{T}
\mathbf{J}_{2} & \cdots & \mathbf{J}_1^\mathsf{T} \mathbf{J}_{N} \\
(N-1)\mathbf{J}_2^\mathsf{T} \mathbf{J}_{1} & (N-1)\mathbf{J}_2^\mathsf{T}
\mathbf{J}_{2} & \cdots &  \mathbf{J}_2^\mathsf{T} \mathbf{J}_{N} \\
\vdots & \vdots & \ddots & \vdots \\
\mathbf{J}_N^\mathsf{T} \mathbf{J}_{1} & \mathbf{J}_N^\mathsf{T} \mathbf{J}_{2}
& \cdots & \mathbf{J}_N^\mathsf{T} \mathbf{J}_{N}
\end{bmatrix} ,
\end{equation*}
where $r_1$-dependence is completely isolated in the first block row and column.
Expanding the block matrices, one can obtain $r_1$-dependence of the matrix
elements:
\begin{equation*}
\mathbf{M}_u = m\begin{bmatrix} 
N & 0 & 0 & ... \\
0 & Nr_1^2 & 0 & \propto r_1 & \cdots & \propto r_1 \\
0 & 0 & Nr_1^2\sin^2\theta_1 & \propto r_1 & \cdots & \propto r_1  \\
\vdots & \propto r_1 & \propto r_1 & \ddots\\
 & \vdots & \vdots \\
 & \propto r_1 & \propto r_1 &
\end{bmatrix} ,
\end{equation*}
where $\propto r_1$ denotes terms proportional to $r_1$. We can obtain a
$2N\times2N$ submatrix $\mathbf{M}_q$ by excluding the first row and column of
$\mathbf{M}_u$:
\begin{equation}
\mathbf{M}_q = m\begin{bmatrix} 
Nr_1^2 & 0 & \propto r_1 & \cdots & \propto r_1 \\
0 & Nr_1^2\sin^2\theta_1 & \propto r_1 & \cdots & \propto r_1  \\
\propto r_1 & \propto r_1 & \ddots\\
\vdots & \vdots \\
\propto r_1 & \propto r_1 &
\end{bmatrix} .
\end{equation}
It is straightforward to show $\text{det}(\mathbf{M}_q) \propto r_1^4$, which
gives rise to an entropic force $2k_BT/r$ upon constraining $r_1=r$. To
demonstrate this, we transform the partition function for the unconstrained
system to the constrained one~\cite{den2013revisiting}. We use the Hamiltonian of
a freely jointed chain without the end constraint to express the partition
function: 
\begin{equation*}
Q_u=\int e^{-\frac{\beta}{2}\mathbf{p}_u^\mathsf{T}
\mathbf{M}_u^{-1}\mathbf{p}_u} d^{2N+1}ud^{2N+1}p_u .
\end{equation*}
$\mathbf{p}_u$ is a vector of $2N+1$ momenta conjugate to $\mathbf{u}$. Delta
functions are introduced to apply the end constraint on $r_1$: 
\begin{equation*}
Q_c(r)=\int e^{-\frac{\beta}{2}\mathbf{p}_u^\mathsf{T}
\mathbf{M}_u^{-1}\mathbf{p}_u}\delta(r_1-r)\delta(\dot{r}_1)
d^{2N+1}ud^{2N+1}p_u ,
\end{equation*}
where $r$-dependence of the constrained partition function is explicitly shown.
The delta functions can be handled by decomposing the mass metric and using the
transformation rule $d^{2N+1}p_u=d^{2N}p_q(Nmd\dot{r_1})$
\begin{align*}
Q_c(r)&=\int e^{-\frac{\beta}{2}\mathbf{p}_q^\mathsf{T}
\mathbf{M}_q^{-1}\mathbf{p}_q}e^{-\frac{\beta}{2}m\dot{r}_1^2}\delta(r_1-r)
\delta(\dot{r}_1)dr_1(Nmd\dot{r}_1)d^{2N}qd^{2N}p_q\\
&=Nm\int e^{-\frac{\beta}{2}\mathbf{p}_q^\mathsf{T}
\mathbf{M}_q^{-1}(r)\mathbf{p}_q}d^{2N}qd^{2N}p_q .
\end{align*}
Integrating with $\mathbf{p}_q$, we can obtain the $r$-dependence of the
partition function:
\begin{equation}
Q_c(r) \propto \int \sqrt{\text{det}(\mathbf{M}_q(r))} d^{2N}q
\end{equation}

\subsection{Calculation of the metric determinant} 

Chain conformations acquired from equilibirum sampling of the conformational
space, without consideration of momentum space, will require a weighting factor
given by the determinant of the metric tensor in the constrained space. The
metric in the constrained space is generally dense, and so calculating the
determinant $g^\alpha$ is computationally intensive.
Fixman~\cite{fixman1974classical} made use of the simple form of the metric
determinant $g$ in the unconstrained space, to find the desired $g^\alpha$ from
the determinant $g^\beta$ in the smaller, orthogonal subspace spanned by the
constrained dimensions.
\begin{equation}
  g = g^\alpha g^\beta \rightarrow g^\alpha = g/g^\beta
\end{equation}
This is further simplified by considering the inverse $\mathbf{H}$ of the metric
$\mathbf{G}^\beta$ for the orthogonal subspace of constrained dimensions. This
has the form
\begin{equation}
 H_{ij} = \sum_{k=0}^N \frac{\partial r_i}{\partial \mathbf{x}_k} \frac{\partial
 r_j}{\partial \mathbf{x}_k}
\end{equation}
where $r_i$ is the bond length between points $\mathbf{x}_{i-1}$ and
$\mathbf{x}_i$.
The terms in the sum are only non-zero when $i = j\pm1$ and $k = i\mbox{ or }
i-1$.
This ultimately produces a tridiagonal matrix 
\begin{equation}
\mathbf{H} = 
\begin{bmatrix}
 2 & -\mathbf{\hat r}_1\cdot \mathbf{\hat r}_{2} & 0 & \cdots & 0 & 0 \\
 -\mathbf{\hat r}_1\cdot \mathbf{\hat r}_{2} & 2 & 
 -\mathbf{\hat r}_2\cdot \mathbf{\hat r}_{3} & \cdots & 0 & 0 \\
 0 & -\mathbf{\hat r}_2\cdot \mathbf{\hat r}_{3} & 2 & \cdots & 0 & 0 \\
 \vdots & \vdots & \vdots & \ddots & \vdots & \vdots \\
 0 &0 & 0 & \cdots & 2 & -\mathbf{\hat r}_{N-1}\cdot \mathbf{\hat r}_{N} \\
 0 & 0 & 0 & \cdots &  -\mathbf{\hat r}_{N-1}\cdot \mathbf{\hat r}_{N} & 2 
\end{bmatrix} .
\end{equation}
Expanding this matrix in minors gives a recursive formula for the determinant
$h$, where $\mathbf{\hat r}_n$ is the unit relative position vector pointing
from (n-1)-th mass point to n-th mass point.
\begin{equation}
 h_n = 2 h_{n-1} - (\mathbf{\hat r}_n\cdot \mathbf{\hat r}_{n-1})^2 h_{n-2}
\end{equation}
For equal mass points, we begin with the base cases $h_0 =h_1 = 1$. Note this is
slightly different than the Fixman case ($h_1 = 2$) as we are also constraining
the position of the first mass point. Once we have computed $h$, the determinant
$g^\alpha$ is easily found from 
\begin{equation}
 g^\alpha = g \cdot h
\end{equation}

\end{document}